\newcommand{\CLASSINPUTtoptextmargin}{0.75in}%
\newcommand{\CLASSINPUTbottomtextmargin}{1in}%
\newcommand{\CLASSINPUTinnersidemargin}{0.63in}%
\newcommand{\CLASSINPUToutersidemargin}{0.63in}%
\documentclass[conference,10pt,letterpaper]{IEEEtran}%
\usepackage{cite}
\usepackage{amsmath}
\usepackage{graphicx}
\usepackage{multirow}
\usepackage[none]{hyphenat}
\usepackage{float}
\usepackage{subfig}
\usepackage{dblfloatfix}

\usepackage{tikz}

\newcommand{\copyrighttext}{%
  \footnotesize \textcopyright 2021 IEEE. DOI: 10.1109/WAMICON47156.2021.9443620 Personal use of this material is permitted.
  Permission from IEEE must be obtained for all other uses, in any current or future
  media, including reprinting/republishing this material for advertising or promotional
  purposes, creating new collective works, for resale or redistribution to servers or
  lists, or reuse of any copyrighted component of this work in other works.
}
\newcommand{\copyrightnotice}{%
\begin{tikzpicture}[remember picture,overlay]
\node[anchor=south,yshift=10pt] at (current page.south) {\fbox{\parbox{\dimexpr\textwidth-\fboxsep-\fboxrule\relax}{\copyrighttext}}};
\end{tikzpicture}%
}

\usepackage{t1enc}
\usepackage{times}
\makeatletter

\def\@IMSauthorblockNAMEstyle{\normalfont\IMSauthorsize}
\def\@IMSauthorblockAFFILstyle{\normalfont\IMSaffilsize}
\def\@IMSauthorblockEMAILstyle{\normalfont\IMSaffilsize}
\def\IMSauthorblockNAME#1{%
\relax\@IMSauthorblockNAMEstyle%
#1%
}%
\def\IMSauthorblockAFFIL#1{%
\relax\@IMSauthorblockAFFILstyle%
\vskip\@IEEEauthorblockAtopspace
#1%
}%
\def\IMSauthorblockEMAIL#1{%
\relax\@IMSauthorblockEMAILstyle%
\vskip\@IEEEauthorblockAtopspace
#1%
}%
\newcommand{\IMSauthor}[1]{%
\ifIsBlindReviewVersion%
\author{\phantom{\parbox{\textwidth}{\center\relax#1}}}%
\else%
\author{\parbox{\textwidth}{\center\relax#1}}%
\fi%
}%
\newif\ifIsBlindReviewVersion

\def\IMSthispaperforfinalpublication{\IsBlindReviewVersionfalse}
\def\@maketitle{\newpage
\bgroup\par\addvspace{0.5\baselineskip}\centering%
\ifCLASSOPTIONtechnote
   {\bfseries\large\@IEEEcompsoconly{\sffamily}\@title\par}\vskip 1.3em{\lineskip .5em\@IEEEcompsoconly{\sffamily}\@author
   \@IEEEspecialpapernotice\par{\@IEEEcompsoconly{\vskip 1.5em\relax
   \@IEEEtitleabstractindextextbox{\@IEEEtitleabstractindextext}\par
   \hfill\@IEEEcompsocdiamondline\hfill\hbox{}\par}}}\relax
\else
   \vskip0.2em{\IMStitlesize\ifCLASSOPTIONtransmag\bfseries\LARGE\fi\@IEEEcompsoconly{\sffamily}\@IEEEcompsocconfonly{\normalfont\normalsize\vskip 2\@IEEEnormalsizeunitybaselineskip
   \bfseries\Large}\@title\par}\vskip1.0em\par
   \ifCLASSOPTIONconference%
      {\@IEEEspecialpapernotice\mbox{}\vskip\@IEEEauthorblockconfadjspace%
       \mbox{}\hfill\begin{@IEEEauthorhalign}\@author\end{@IEEEauthorhalign}\hfill\mbox{}\par}\relax
   \else
      \ifCLASSOPTIONpeerreviewca
         {\@IEEEcompsoconly{\sffamily}\@IEEEspecialpapernotice\mbox{}\vskip\@IEEEauthorblockconfadjspace%
          \mbox{}\hfill\begin{@IEEEauthorhalign}\@author\end{@IEEEauthorhalign}\hfill\mbox{}\par
          {\@IEEEcompsoconly{\vskip 1.5em\relax
           \@IEEEtitleabstractindextextbox{\@IEEEtitleabstractindextext}\par\hfill
           \@IEEEcompsocdiamondline\hfill\hbox{}\par}}}\relax
      \else
         \ifCLASSOPTIONtransmag
           {\@IEEEspecialpapernotice\mbox{}\vskip\@IEEEauthorblockconfadjspace%
            \mbox{}\hfill\begin{@IEEEauthorhalign}\@author\end{@IEEEauthorhalign}\hfill\mbox{}\par
           {\vspace{0.5\baselineskip}\relax\@IEEEtitleabstractindextextbox{\@IEEEtitleabstractindextext}\vspace{-1\baselineskip}\par}}\relax
         \else
           {\lineskip.5em\@IEEEcompsoconly{\sffamily}\sublargesize\@author\@IEEEspecialpapernotice\par
           {\@IEEEcompsoconly{\vskip 1.5em\relax
            \@IEEEtitleabstractindextextbox{\@IEEEtitleabstractindextext}\par\hfill
            \@IEEEcompsocdiamondline\hfill\hbox{}\par}}}\relax
         \fi
      \fi
   \fi
\fi\par\addvspace{0.0\baselineskip}\egroup}

\def\IMStitlesize{\@setfontsize{\IMStitlesize}{18}{21pt}}
\def\IMSauthorsize{\@setfontsize{\IMSauthorsize}{12}{13pt}}
\def\IMSaffilsize{\@setfontsize{\IMSaffilsize}{12}{13pt}}
\def\IMScaptionsize{\@setfontsize{\IMScaptionsize}{8}{9pt}}
\def\IMSbibsize{\@setfontsize{\IMSbibsize}{8}{9pt}}

\def\@IEEEauthorblockNstyle{\IMSauthorsize\@IEEEcompsocnotconfonly{\sffamily}\@IEEEcompsocconfonly{\large}}
\def\@IEEEauthorblockAstyle{\IMSaffilsize\@IEEEcompsocnotconfonly{\sffamily}\@IEEEcompsocconfonly{\itshape}\@IEEEcompsocconfonly{\large}}
\def\@IEEEauthordefaulttextstyle{\IMSauthorsize\@IEEEcompsocnotconfonly{\sffamily}\sublargesize}

\def\thebibliography#1{\section*{\refname}%
    \addcontentsline{toc}{section}{\refname}%
    \IMSbibsize\@IEEEcompsocconfonly{\small}\vskip 0.3\baselineskip plus 0.1\baselineskip minus 0.1\baselineskip
    \list{\@biblabel{\@arabic\c@enumiv}}%
    {\settowidth\labelwidth{\@biblabel{#1}}%
    \leftmargin\labelwidth
    \advance\leftmargin\labelsep\relax
    \itemsep \IEEEbibitemsep\relax
    \usecounter{enumiv}%
    \let\p@enumiv\@empty
    \renewcommand\theenumiv{\@arabic\c@enumiv}}%
    \let\@IEEElatexbibitem\bibitem%
    \def\bibitem{\@IEEEbibitemprefix\@IEEElatexbibitem}%
\def\newblock{\hskip .11em plus .33em minus .07em}%
\ifCLASSOPTIONtechnote\sloppy\clubpenalty4000\widowpenalty4000\interlinepenalty100%
\else\sloppy\clubpenalty4000\widowpenalty4000\interlinepenalty500\fi%
    \sfcode`\.=1000\relax}

\long\def\@makecaption#1#2{%
\ifx\@captype\@IEEEtablestring%
\par\@IEEEtabletopskipstrut
\else
\@IEEEfigurecaptionsepspace
\fi
\setbox\@tempboxa\hbox{\normalfont\IMScaptionsize {#1.}\nobreakspace\nobreakspace #2}%
\ifdim \wd\@tempboxa >\hsize%
\setbox\@tempboxa\hbox{\normalfont\IMScaptionsize {#1.}\nobreakspace\nobreakspace}%
\parbox[t]{\hsize}{\normalfont\IMScaptionsize\noindent\unhbox\@tempboxa#2}%
\else
\ifCLASSOPTIONconference \hbox to\hsize{\normalfont\IMScaptionsize\hfil\box\@tempboxa\hfil}%
\else \hbox to\hsize{\normalfont\IMScaptionsize\box\@tempboxa\hfil}%
\fi\fi
\ifx\@captype\@IEEEtablestring%
\@IEEEtablecaptionsepspace
\else
\fi}

\newlength\tablecaptiontotableskip
\newlength\figuretocaptionskip
\def\@IEEEfigurecaptionsepspace{\vskip\figuretocaptionskip\relax}%
\def\@IEEEtablecaptionsepspace{\vskip\tablecaptiontotableskip\relax}%

\def\abstract{\normalfont%
\@IEEEabskeysecsize\bfseries\textit{\abstractname}\,\bfseries\textit{---}\,%
\@IEEEgobbleleadPARNLSP}%

\def\IEEEkeywords{\normalfont%
\@IEEEabskeysecsize\bfseries\textit{\IEEEkeywordsname}\,\bfseries\textit{---}\,%
\@IEEEgobbleleadPARNLSP}%
\def\endIEEEkeywords{\relax\vspace{0.67ex}%
\par\if@twocolumn\else\endquotation\fi%
\normalsize\normalfont}%

\DeclareRobustCommand*{\IMSauthorrefmark}[1]{\raisebox{0pt}[0pt][0pt]{\textsuperscript{\footnotesize{#1}}}}%
\def\@IEEEauthorblockNtopspace{0ex}
\def\@IEEEauthorblockAtopspace{1mm}
\def\tablename{Table}
\def\thetable{\arabic{table}}
\def\IEEEkeywordsname{Keywords}
\def\subsubsection{\@startsection{subsubsection}{3}{\z@}{1.5ex plus 1.5ex minus 0.5ex}%
{0.7ex plus .5ex minus 0ex}{\normalfont\normalsize\itshape}}%
\def\@seccntformat#1{\csname the#1dis\endcsname\relax}
\def\thesectiondis{\thesection.\hskip 0.5em}
\def\thesubsectiondis{{\hbox to\parindent{\Alph{subsection}.}}}
\def\thesubsubsectiondis{{\hbox to \parindent{\arabic{subsubsection})}}}
\def\theparagraphdis{{\hbox to \parindent{\alph{paragraph})}}}
\IEEEilabelindentA \parindent
\IEEEilabelindent \IEEEilabelindentA
\IEEEelabelindent \parindent
\IEEEdlabelindent \parindent
\IEEElabelindent \parindent

\newlength\@IMSparindent
\newcommand\IMSdisplayacksection[1]{%
\ifIsBlindReviewVersion%
\noindent\phantom{\parbox[t]{\columnwidth}{\normalbaselines\setlength{\parindent}{\@IMSparindent}{#1}\strut}}
\else%
\noindent\parbox[t]{\columnwidth}{\normalbaselines\setlength{\parindent}{\@IMSparindent}{#1}\strut}%
\fi%
}%

\makeatother

\begin{document}
\raggedbottom
%
%
%

\title{A Low-Loss 1-4 GHz Optically-Controlled Silicon Plasma Switch}
%
%
%
\IMSthispaperforfinalpublication

\IMSauthor{%
\IMSauthorblockNAME{
Alden Fisher\IMSauthorrefmark{\#1},
Zach Vander Missen\IMSauthorrefmark{\#2},
Abbas Semnani\IMSauthorrefmark{*3},
Dimitrios Peroulis\IMSauthorrefmark{\#4}
}
\\%
\IMSauthorblockAFFIL{
\IMSauthorrefmark{\#}ARES Lab, Purdue University, USA\\
\IMSauthorrefmark{*}University of Toledo, USA
}
\\%
\IMSauthorblockEMAIL{
\IMSauthorrefmark{1}fishe128@purdue.edu, \IMSauthorrefmark{2}zvm@purdue.edu, \IMSauthorrefmark{3}abbas.semnani@utoledo.edu, \IMSauthorrefmark{4}dperouli@purdue.edu
}
}
%
\maketitle
\copyrightnotice
%
%
%
\begin{abstract}
This paper presents a low-loss optically-controlled inline RF switch suitable for L- and S-band applications. Under 1.5 W laser power, the switch exhibits a measured ON-state insertion loss of less than 0.33~dB and return loss better than 20~dB across the band. The measured OFF-state isolation ranges from 27 dB at 1 GHz to 17 dB at 4 GHz. The switch comprises a single silicon chiplet excited by a 915-nm laser fiber which creates electron-hole pairs, thereby exciting the ON-state silicon plasma. An optical fiber is guided through the bottom of the RF substrate to illuminate the chiplet, which bridges a 1.075-mm microstrip line gap. To the best of our knowledge, this is the lowest-loss silicon plasma switch demonstrated today.
\end{abstract}
%
%

\section{Introduction}
Optically-tunable RF circuits offer a variety of attributes, such as fast switching speeds and tunability, without moving mechanical parts. Additionally, these switches provide high power handling at a low production cost when compared to conventional RF switching technologies. 

Current literature discusses utilizing silicon plasma as switches for a variety of applications \cite{Pang2018,Pang2018_1,Tripon-Canseliet2012,Platte2010,Kowalczuk2010,Karabegovic2009,Liu2007,Canseliet2003,Arnould2002,Khalil1997,Gevorgian1994}. Researchers in \cite{Pang2018,Pang2018_1,Kowalczuk2010} focus on the linearity, showing measured IIP3 as high as +77~dBm. Although the reported results primarily focus on switches and applications thereof (e.g.~\cite{Liu2007} and \cite{Arnould2002,Safwat1996} focus on antennas and matching networks respectively), \cite{Mital2008,Ren2018,Jiang2017} have used similar silicon plasma technology to realize variable attenuators.

This technology requires the inclusion of laser diodes to control the plasma state (ON or OFF) in the semiconductor chiplet. This biasing network is relatively straightforward, as it is naturally decoupled from the direct RF path. A relatively recent application domain of this technology is in 5G base stations since these systems require 1) high power handling, 2) fast switching speeds, 3) low cost, and 4) small footprint switching solutions \cite{Pang2018}, \cite{Kowalczuk2010}.

This work improves upon the current state of the art by demonstrating the lowest insertion loss switch in the L and S bands \cite{Pang2018_1}. Additionally, it demonstrates this technology as an SMD replacement for current attenuators, due to the continuously varying response to incident laser power. Moreover, it investigates RF performance improvements achievable by extending power capabilities of the laser bias.



\section{Theory}
When a high-energy photon impinges upon a semiconductor, an electron-hole pair is produced. The accepted bandgap of silicon is 1.12 eV at room temperature, meaning incident light with a wavelength shorter than about one micron can initiate such a phenomenon. However, empirical data in \cite{Mital2008} shows that longer wavelengths, which still create an electron-hole pair, can provide greater penetration into the silicon.

The closed-form expression for exited carriers generated as a function of depth, z, into the sample is given originally by \cite{Platte2010} and improved upon by \cite{KowalczukThesis} to account for higher radiant fluxes ($n > 10^{16} $ $cm^{-3}$)

\begin{figure}[t]
	\centering
	\includegraphics[width=80mm]{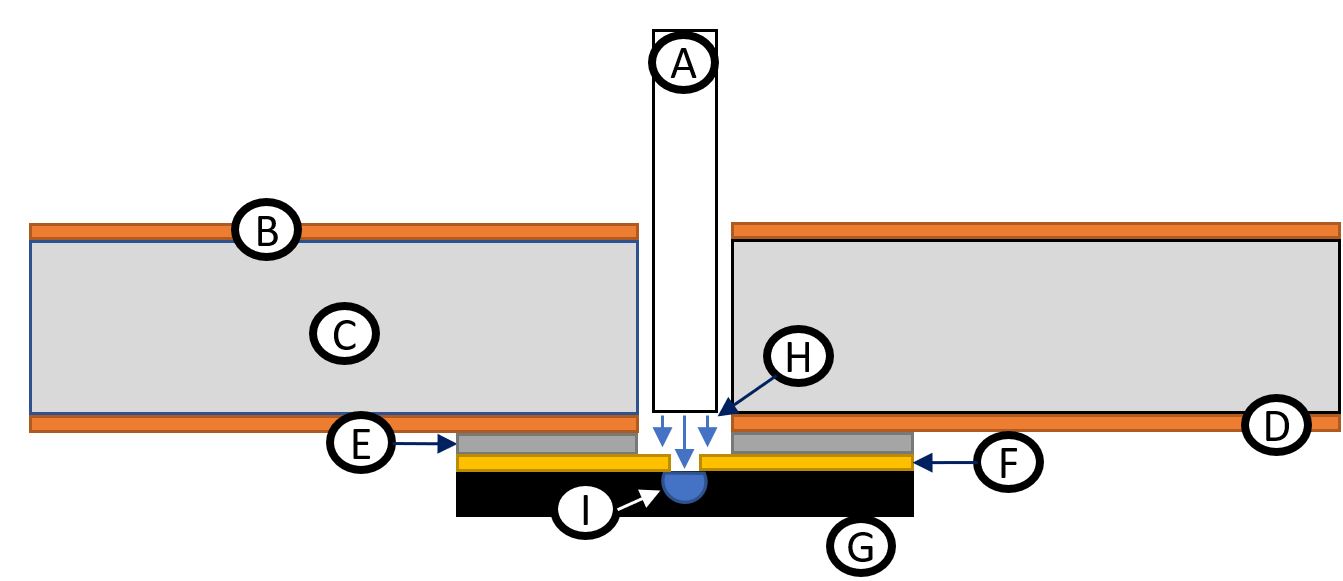}
	\caption{Proposed cross-sectional stack up of inverted board with laser excitation where (A) is the fiber, (B) is the bottom copper, (C) is the TMM3 substrate, (D) is the top copper, (E) is the silver epoxy, (F) is the gold layer, (G) is the silicon die, (H) shows the light illumination path, and (I) shows the free carriers generated.}
	\label{fig:stackup}
\end{figure}

\begin{equation}
\label{eqn:cond}
n(z) = \eta\alpha\tau\frac{P\lambda}{Ahc}\frac{1-R}{1-\alpha^2 L^2}\left[e^{-\alpha z}-\frac{\alpha L^2+\nu_s \tau}{L+\nu_s \tau}e^{-z/L}\right]
\end{equation}

where $\eta$ is the quantum efficiency, $\alpha$ is the absorption coefficient, $\tau$ is the majority carrier lifetime, $P$ is the power of the laser, $\lambda$ is the incident laser light's wavelength, $A$ is the incident area, $R$ is the reflection from the surface, $L$ is the diffusion length, and $\nu_s$ is the saturation velocity at the surface. Certain wavelengths around ~800-1000 nm emperically have the best quantum efficiency, lending to  a higher probability that all the photon packets will be absorbed and create electron-hole pairs. The reflection remains constant around 0.3 for all wavelengths in this range.

\section{Design}

The design consists of a simple microstrip line with a small through gap that has been optimized for the 1 to 4 GHz frequency range. A silicon chiplet is assembled over a small gap in the microstrip line. The chiplet is plated with gold on one side with a small centered gap. This gap is crucial as it exposes the silicon such that laser illumination may be introduced by a fiber which is inserted from the bottom of the RF substrate. As a result, this induces the silicon plasma and creates conduction. A more detailed view may be seen in Fig.~\ref{fig:stackup}.

In order to optimize the ON/OFF ratio, a 200~$\mu$m lightly doped n-type silicon wafer is selected to manufacture the chiplet. With a phosphorous doping concentration of about $1.5\times10^{12}$ $cm^{-3}$ (3000~$\Omega\times cm$), more free carriers are initially present, when compared to intrinsic silicon, allowing better insertion loss. This preserves a high degree of isolation, which increases the ON/OFF ratio of the switch. A 915-nm wavelength laser is used, as this provides both high quantum efficiency and deep penetration into the chiplet, allowing better contact with the gold. The laser output power varies continuously up to 1.5~W, ensuring that the achievable conductivity has a wide range.

The silicon chiplets are sputtered with 2~$\mu$m of gold on one side and measure 3.075~mm by 500~$\mu$m by 200~$\mu$m (length by width by height) with a 75-$\mu$m gap in the gold in the middle. This gap allows the 100~$\mu$m fiber to illuminate the exposed silicon, permitting the two sides to conduct.

The chiplets are custom manufactured in-house. First, the 4-inch bare die is sputtered with titanium to form a coating 20~nm in thickness, allowing good adhesion to the semiconductor. Thereafter, a 2~$\mu$m gold layer is sputtered on top. After lithography, the 75-$\mu$m gap is etched of both gold and titanium. A dicing saw is then used to create the aforementioned sizes, which are comparable to current SMDs.

\begin{figure}
	\centering
	\includegraphics[width=75mm]{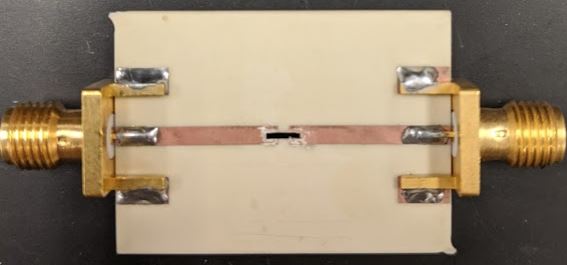}
	\caption{Evaluation board with silicon chiplet mounted.}
	\label{fig:board}
\end{figure}

The RF board is a 30 mil TMM3 ($\epsilon_r = 3.45$) substrate with 17.5 $\mu$m of copper thickness. The top pattern is a simple 50-$\Omega$ microstrip through with a 1.075 mm gap in the middle. A via is drilled in the center of this gap for the fiber. The final board is shown in Fig. \ref{fig:board}.

The chiplet is assembled over and contacted to the microstrip line using silver epoxy. A 200~$\mu$m non-plated via, directly under the 75~$\mu$m exposed silicon, allows the fiber to be inserted through the RF substrate. The stack up depicted in Fig.~\ref{fig:stackup} is comparable to \cite{Pang2018,Pang2018_1}, except a micromanipulator (which can be seen in Fig.~\ref{fig:setup}) is used to guide the fiber through the via. Furthermore, the chiplet is attached to the board much like an SMD, rather than compressed against the board and capacitavely coupled. This contributes to improved RF performance. Note that the RF board is technically upside-down during optical excitation to allow the micromanipulator to position the fiber through the via on the bottom side of the board. While testing, the board is raised from the table to reduce any effects from materials other than free space.

\section{Measurements}

\begin{figure} [h!]
	\centering
	\includegraphics[width=80mm]{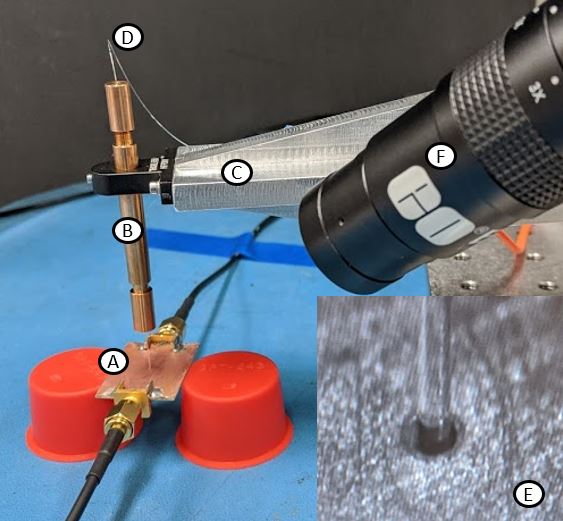}
	\caption{The measurement setup. The board (A) is raised above the table so that the fiber positioning with the fiber chuck (B) can take place while not causing RF interference. The micromanupulator (C) positioning the fiber (D) on screen is shown in the inset (E) via the camera (F).}
	\label{fig:setup}
\end{figure}


\begin{figure}[h!]
	\centering%
	\subfloat[]{%
		\centering
		\includegraphics[width=80mm]{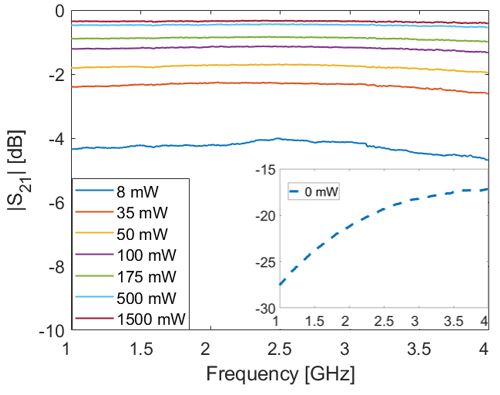}
		\label{fig:IL}
	}%
	\\[2.6mm]
	\subfloat[]{%
		\centering
		\includegraphics[width=80mm]{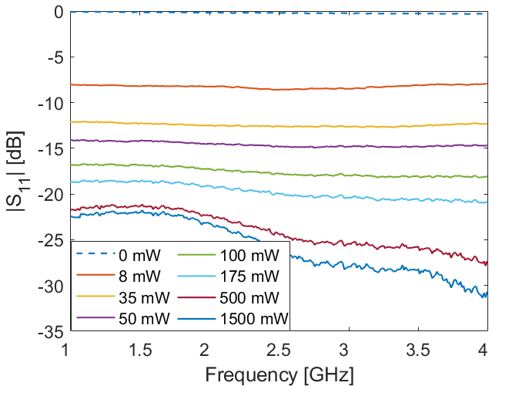}
		\label{fig:RL}
	}%
	\\[-0.1mm]%
	\caption{Measured S-parameters of the switch's ON- and OFF-state. (a) insertion loss at several laser powers; (b) return loss at several laser powers.}
\end{figure}

\begin{figure}[h!]
	\centering%
	\subfloat[]{%
		\centering
		\includegraphics[width=35mm]{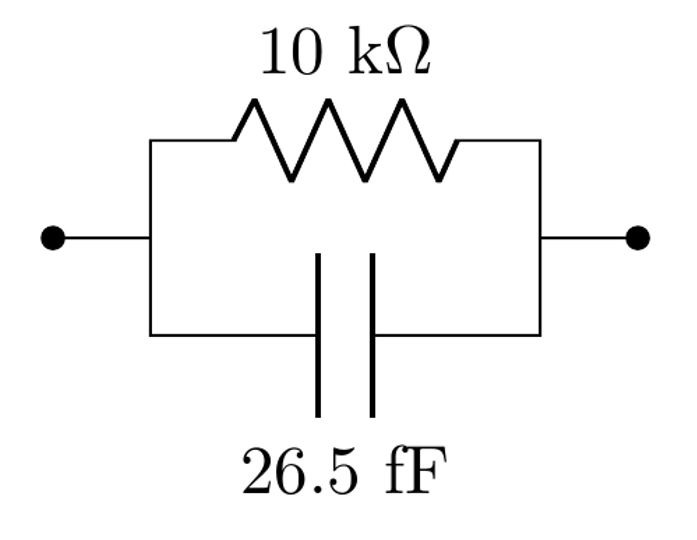}
		\label{fig:off}
	}%
	\\[-0.1mm]%
	\subfloat[]{%
		\centering
		\includegraphics[width=35mm]{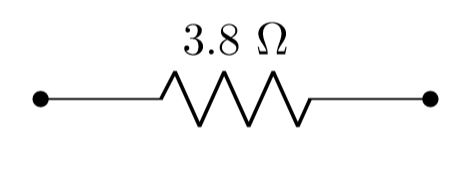}
		\label{fig:on}
	}%
	\\[-0.1mm]%
	\caption{Extracted simplified equivalent circuit for the switch. (a) OFF state (0 mW laser excitation); (b) ON state (1.5 W laser excitation).}
\end{figure}

The set up consists of a 1.5-W, quasi-CW fiber-coupled laser from Sheaumann lasers centered at 915 nm. The multimode fiber has a nominal diameter of 100 $\mu$m and is cleaved at the end to allow for free space coupling. The fiber is held by a fiber chuck which is affixed to a cantilevered arm, connected to a three-axis micromanipulator for precise positioning of the fiber's tip. In the set up, shown in Fig.~\ref{fig:setup}, a camera is focused on the location of the via, displaying the fiber's position. Once the fiber is in the via, it is lowered until contact is made with the silicon chiplet, then raised by several microns. This allows the light to disperse which increases the contact area on the silicon, forming better contact between the two gold sides of the chiplet.

The board is excited under nominal RF power while varying the laser power. Constant closed-loop monitoring is used to control the diode's junction temperature to be within 10$^\circ$ C of ambient. The laser power is varied from 0 mW up to 1.5 W.

Measured results from the switch are shown in Figs.~\ref{fig:IL} and \ref{fig:RL}. Under no laser illumination (i.e. 0 mW) the isolation is greater than 17 dB and the return loss is less than 0.30 dB for the 1 to 4 GHz range. Upon illumination, the chiplet approximates a short circuit. Note that this trend is continuously varying, allowing the chiplet to act as a variable attenuator controlled by the illumination power.

\subsection{Comparison with State of the Art}
Table \ref{tab:SOA} shows the current literature focused on creating silicon plasma switches. It is important to note that the advanced results achieved in this work are not solely attributed to increases in power, as results surpassing the state of the art are achieved at similar powers.

\begin{table}
	\caption{This work compared to published measured silicon plasma switches.}
	\small
	\centering
	\begin{tabular}{|c|c|c|c|}\hline
		\multirow{1}{6mm}{\parbox{8mm}{{\bfseries Ref.}}} & \raisebox{-0.25mm}{\bfseries Freq. [GHz]} & \raisebox{-0.25mm}{\bfseries Laser Power [mW]} & \raisebox{-0.25mm}{\bfseries Loss [dB]} \\ \hline
		
		\raisebox{-0.25mm}{\cite{Pang2018_1}} & \raisebox{-0.25mm}{32-50} & \raisebox{-0.25mm}{175} & \raisebox{-0.25mm}{<4}\\ \hline
		
		\raisebox{-0.25mm}{\cite{Pang2018}} & \raisebox{-0.25mm}{0-6} & \raisebox{-0.25mm}{175} & \raisebox{-0.25mm}{1.11} \\ \hline
		
		\raisebox{-0.25mm}{This Work} & \raisebox{-0.25mm}{1-4} & \raisebox{-0.25mm}{175}&\raisebox{-0.25mm}{0.84} \\ 
		
		\hline
		
		\raisebox{-0.25mm}{\cite{Kowalczuk2010}} & \raisebox{-0.25mm}{1-6}& \raisebox{-0.25mm}{200}& \raisebox{-0.25mm}{0.69}\\ \hline
		
		\raisebox{-0.25mm}{This Work} & \raisebox{-0.25mm}{1-4} & \raisebox{-0.25mm}{200}&\raisebox{-0.25mm}{0.72} \\ \hline
		
		\raisebox{-0.25mm}{This Work} & \raisebox{-0.25mm}{1-4} & \raisebox{-0.25mm}{1500}&\raisebox{-0.25mm}{0.33} \\ \hline
	\end{tabular}
	\label{tab:SOA}
\end{table}

\subsection{Circuit Model}

In Fig.~\ref{fig:off}, a simplified equivalent circuit model for the switch's OFF state is seen. The resistor, which dominates the performance at lower frequencies, is due to the bulk resistivity of the doped die and its skin depth. The parallel capacitance is due to the parasitic capacitance between the gold traces on the chiplet. Figure~\ref{fig:on} gives the simplified equivalent circuit for the switch in the ON state. As it is simply a resistor, this indicates that the response is flat, at least for lower frequencies. It should be noted here that the resistor value is a function of the laser power.


{
\setlength{\tabcolsep}{1mm}%
\newcommand{\CPcolumnonewidth}{50mm}%
\newcommand{\CPcolumntwowidth}{91mm}%
\newcommand{\CPcell}[1]{\hspace{0mm}\rule[-0.3em]{0mm}{1.3em}#1}%
\newcommand{\CPcellbox}[1]{\parbox{90mm}{\hspace{0mm}\rule[-0.3em]{0mm}{1.3em}#1\strut}}%

}



\section{Conclusion}

An optically-controlled RF switch has been demonstrated with state-of-the-art results in the 1 to 4~GHz bandwidth. The OFF-state isolation is better than 17~dB and the ON-state insertion loss is better than 0.33~dB. The ON-state return loss is better than 22~dB across the 4-to-1 band under 915 nm, 1.5 W laser excitation. The proposed switch is a viable replacement for many conventional switches -- and by extension -- antenna arrays, impedance tuners, filters, among others. The reported performance has been achieved by a combination of chiplet design, mounting techniques, and illumination power. This work shows the viability of silicon plasma technology as an RF switch.


\section*{Acknowledgment}


\newcommand{\IMSacktext}{%
	The authors would like to acknowledge the extensive help from Michael Sinanis who helped with the wafer/chiplet fabrication.
	
	This effort undertaken was sponsored by the Department of
	the Navy, Office of Naval Research under ONR award number
	N00014-19-1-2549.  This  work  relates  to  Department  of
	Navy   award   N00014-19-1-2549   issued   by   the   Office
	of  Naval  Research.  The  United  States  Government  has  a
	royalty-free license throughout the world in all copyrightable
	material   contained   herein.   Any   opinions,   findings,   and
	conclusions or recommendations expressed in this material are
	those of the authors and do not necessarily reflect the views
	of the Office of Naval Research.
}

\IMSdisplayacksection{\IMSacktext}


\bibliographystyle{IEEEtran}

\bibliography{IEEEexample}

\end{document}